\documentclass[aps,prd,reprint,amsmath,amssymb,superscriptaddress,longbibliography,nofootinbib,floatfix]{revtex4-2}
\usepackage[T1]{fontenc}
\usepackage{lmodern}
\input{glyphtounicode}\pdfgentounicode=1
\usepackage{graphicx,bm,booktabs,mathrsfs,etoolbox,array}
\usepackage[colorlinks=true,citecolor=blue,linkcolor=blue,urlcolor=blue]{hyperref}
\graphicspath{{figures/}}
\usepackage{mathtools}

\begin{document}
%\title{Novel inverse-temperature asymptotics in black hole thermodynamic topology}
% \title{Novel inverse-temperature asymptotics in black hole thermodynamic topology}
\title{Novel thermodynamic topological subclass from inverse-temperature asymptotics}
\author{Dejiang Yin}
\email{yyin.dejiang@gmail.com}
\affiliation{College of Physics, Guizhou University, Guiyang 550025, People’s Republic of China}

\author{Qi-Qi Liang}
%\email{yyin.dejiang@gmial.com}
\affiliation{College of Physics, Guizhou University, Guiyang 550025, People’s Republic of China}

\author{Yu-Die Wan}
%\email{yyin.dejiang@gmial.com}
\affiliation{College of Physics, Guizhou University, Guiyang 550025, People’s Republic of China}

\author{Li-yun Zhang}
% \email{liy\_zhang@hotmail.com (Corresponding author)}
\email{liy\_zhang@hotmail.com}
\affiliation{College of Physics, Guizhou University, Guiyang 550025, People’s Republic of China}

% \date{August 27, 2026}
% \date{\today}

% \email[Corresponding author: ]{liy\_zhang@hotmail.com,yyin.dejiang@gmail.com}

% \date{October 11, 2026}
\begin{abstract}

We investigate the thermodynamic topology of the charged
McGuigan--Nappi--Yost black hole in the canonical ensemble
at fixed electric charge,
identifying equilibrium black hole solutions as zeros
of the thermodynamic vector field constructed from
the generalized off-shell free energy.
The inverse Hawking temperature diverges as the extremal
limit is approached and approaches a finite positive
value in the large-horizon limit. The zero points of the
thermodynamic vector field reveal a thermodynamic phase
structure consisting solely of a locally stable branch
with positive winding number ($w=+1$).
The innermost small black hole state is locally stable
in the low-temperature limit. As the temperature
increases, the stable branch extends toward the
outermost large black hole regime, while no black hole
states exist in the high-temperature regime beyond the
finite upper Hawking temperature.
This distinct asymptotic behavior of the inverse
temperature places the thermodynamic defect structure
outside the established topological (sub)classes,
although its topological number agrees with that of
the standard $W^{1+}$ class.
We therefore identify this defect structure as a new
thermodynamic topological subclass, $\breve W^{1+}$.
This finding extends the existing classification to include
a finite, nonzero large-horizon limit of the inverse Hawking
temperature and suggests that distinct asymptotic behaviors
in other black hole families may reveal further topological
(sub)classes and thermodynamic phase structures.

\end{abstract}\maketitle

\section{Introduction}

Black hole thermodynamics relates horizon mechanics, entropy, and Hawking radiation~\cite{1973CMaPh..31..161B,1973PhRvD...7.2333B,1974Natur.248...30H,1975CMaPh..43..199H}. Euclidean actions and boundary energies identify the thermodynamic potentials appropriate to specified boundary data~\cite{1977PhRvD..15.2752G,1986PhRvD..33.2092Y,1993PhRvD..47.1407B}. In charged systems the choice between fixed charge and fixed potential changes the relevant free energy and can change the phase structure~\cite{1999PhRvD..60f4018C,1999PhRvD..60j4026C}.
This thermodynamic framework also provides a basis for
characterizing black-hole equilibrium states from a
topological perspective.

In this approach, black-hole equilibrium states are identified as zeros of a vector field constructed from an
appropriate generalized off-shell thermodynamic potential
\cite{2022PhRvL.129s1101W}. In the canonical ensemble
considered here, this potential takes the Helmholtz form,
with an auxiliary inverse temperature that equals the
inverse Hawking temperature at equilibrium. Duan's $\phi$-mapping
topological current theory assigns a local winding number
to each isolated zero
\cite{Duan:1984ws,1998NuPhB.514..705D}. Positive and
negative winding numbers distinguish locally stable and
unstable states, respectively. Wei et al. classified these
states using their topological numbers, limiting branches,
and low- and high-temperature behavior
\cite{2024PhRvD.110h1501W}. Later work further examined
the generation and annihilation of defect pairs
\cite{2025PhRvD.111f1501W}. The topology of equilibrium black-hole states complements
that of thermodynamic critical points
\cite{2022PhRvD.105j4003W,2022PhRvD.105j4053Y};
both approaches are reviewed in
Refs.~\cite{2025EPJC...85.1372W,2026SCPMA..6960401W}.

The thermodynamic topology framework has been extended to a
wide range of gravitational systems, where different physical
ingredients modify the structure of thermodynamic defects.
Rotating and accelerating black holes can alter the number of
branches and their asymptotic behaviors
\cite{2023PhRvD.107b4024W,2023PhRvD.107h4002W,
2023PhRvD.108h4041W,2025CQGra..42l5007L}.
Charged Taub--NUT and de Sitter black holes further emphasize
the importance of consistent thermodynamic variables and
physical horizon domains
\cite{2023EPJC...83..589W,2023EPJC...83..927D}. 
Thermodynamic topology has also been studied for charged dilatonic black holes in different ensembles, with branch structures and global topological numbers depending on the dilaton coupling, horizon curvature, and cosmological constant \cite{2024EPJC...84.1204H}.
Ensemble choices, such as those in
dyonic systems, and modifications of the entropy prescription
can also change the corresponding off-shell thermodynamic
field
\cite{2023PhRvD.108f6016G,2024PhLB..85638885Z,
2024EPJC...84..826T,2025PhLB..86839720L}.
Moreover, restricted or extended phase-space descriptions,
BTZ and matter-coupled black holes, as well as finite-cavity
systems, demonstrate that the physical parameter domain and
boundary conditions play an essential role in determining the
defect curves and their global topological structures
\cite{2024AnPhy.46069569S,2024Symm...16.1577D,
2025PhLB..86539482C,2023EPJC...83..944R,
2025EPJC...85.1322Z,2026EPJC...86..929Z}.

Thermodynamic topological (sub)classes are characterized by
the global topological number $W$, the ordered sequence of
local winding numbers, and the asymptotic behavior of the
inverse temperature $\beta(r_h)$ at the boundaries of the
physical horizon domain.
The evolution of defect branches, including generation
and annihilation points when present, provides further
information for distinguishing subclasses~\cite{2024PhRvD.110h1501W,2025PhRvD.111f1501W,2026SCPMA..6960401W}.
The four original classes are $W^{1-}$, $W^{0+}$,
$W^{0-}$, and $W^{1+}$, distinguished by the
inverse-temperature asymptotics as $r_h\to r_m^+$
and $r_h\to\infty$:
$\beta(r_m),\beta(\infty)\in\{0,\infty\}$
\cite{2024PhRvD.110h1501W}.
Subsequent studies extended this classification to
additional classes and subclasses with distinct branch
structures and inverse-temperature asymptotics:
$W^{0-\leftrightarrow1+}$, $\overline W^{1+}$,
$\widehat W^{1+}$, $\widetilde W^{1+}$,
$\ddot W^{1-}$, $\overline W^{0-}$,
$\widetilde W^{0-}$, and $\overline W^{1-}$
\cite{2025PhRvD.111f1501W,2025PhRvD.112l4024A,
2025EPJC...85.1386C,2026EPJC...86..187W,
2026arXiv260714173W,2026arXiv260916568Z,
2026arXiv260920898Y}.
Wu et al. identified $W^{0-\leftrightarrow1+}$ and
$\overline W^{1+}$ subclasses, for which $\beta(r_m)$ approaches
a finite positive value (termed ``fixed temperature'' in
Ref.~\cite{2025PhRvD.111f1501W}), while
$\beta(\infty)=0$. The same study conjectured
$W^{0+\leftrightarrow1-}$, $\overline W^{1-}$,
and $\widehat W^{1-}$ from the symmetry of the
$(r_h,\beta)$ profiles. The first two have a finite
positive $\beta(r_m)$ and $\beta(\infty)=\infty$,
whereas $\widehat W^{1-}$ has
$\beta(r_m)=0$ and $\beta(\infty)=\infty$
\cite{2025PhRvD.111f1501W}.
The $\overline W^{1-}$ subclass was recently realized
in black-hole solutions
\cite{2026arXiv260916568Z,2026arXiv260920898Y}.
Despite these extensions, none of the established or conjectured (sub)classes listed above has a finite, nonzero inverse-temperature limit as $r_h\to\infty$, that is, $\beta(\infty)=\beta_{\mathrm f}\in(0,\infty)$ (or $\beta(\infty)=$fixed temperature; see Ref.~\cite{2025PhRvD.111f1501W}).

Two-dimensional dilaton gravity provides an analytically
tractable setting in which black-hole entropy is determined
by the dilaton rather than a horizon area
\cite{2002PhR...369..327G,2007JHEP...04..074G,
1993CQGra..10.1477M,1996PhRvD..54.7476C}.
Earlier studies developed string black-hole geometries
\cite{1991PhRvD..44..314W,1992PhRvD..45.1005C,
1992NuPhB.371..269D} and charged solutions in
two-dimensional dilaton gravity
\cite{1991PhRvL..67.3336I,1995PhRvD..52.3494L}.
McGuigan et al. constructed a charged heterotic
black hole with inner and outer horizons
\cite{1992NuPhB.375..421M}. Nappi \& Pasquinucci
subsequently derived its thermodynamic quantities
\cite{1992MPLA....7.3337N}. This solution thus provides
an analytic setting for examining the large-horizon
asymptotic behavior of the inverse temperature.

In this work, we examine the thermodynamic topology of
the charged McGuigan--Nappi--Yost (MNY) black hole in the
canonical ensemble. We derive the equilibrium defect curve
from the generalized off-shell free energy and determine
the winding number and boundary behavior of its black-hole
states. We find a single stable branch with positive
winding number, for which the inverse Hawking temperature
approaches a finite, nonzero value as $r_h\to\infty$.
This asymptotic behavior distinguishes it from the
standard $W^{1+}$ class, where $\beta(\infty)=0$
\cite{2024PhRvD.110h1501W,2025PhRvD.111f1501W}.
We therefore identify this defect structure as a new
thermodynamic topological subclass, $\breve W^{1+}$.

The remainder of this paper is organized as follows.
In Sec.~\ref{MNY:solution}, we review the charged MNY solution and its
thermodynamic quantities. In Sec.~\ref{MNY:off-shell}, we construct the
off-shell vector field and analyze its zeros, winding
numbers, and boundary behavior. In Sec.~\ref{MNY:topo_class}, we compare
the resulting defect structure with existing thermodynamic
topological classifications. Section~\ref{MNY:conclusion} summarizes our
results.

\section{Charged MNY solution and thermodynamics}
\label{MNY:solution}

We consider the charged heterotic black hole constructed by
McGuigan et al.~\cite{1992NuPhB.375..421M}, using the
notation and action normalization of Nappi and
Pasquinucci~\cite{1992MPLA....7.3337N}. With metric
signature $(-,+)$ and in units $\hbar=c=k_B=1$, the
solution is
\begin{equation}
\begin{aligned}
ds^2&=-g(x)dt^2+g(x)^{-1}dx^2,\\
g(x)&=1-2me^{-Qx}+q^2e^{-2Qx},\\
\phi(x)&=\phi_0-\tfrac12 Qx,\\
F&=f(x)\,dx\wedge dt,\qquad f(x)=\sqrt2 Qq e^{-Qx}.
\end{aligned}\label{metric}
\end{equation}
Here $\phi$ is the dilaton, $\phi_0$ is a fixed real
constant, and $Q>0$ has dimensions of inverse length.
The dimensionless parameters $m$ and $q$ determine
the mass and electric charge, respectively.
We take $m>|q|>0$ for the nonextremal charged solution.
The condition $g(\infty)=1$ fixes the asymptotic time
normalization. The gauge potential is
$A_t(x)=-\sqrt2q e^{-Qx}$, with $A_t(\infty)=0$.
The bulk action is
\cite{1992NuPhB.375..421M,1992MPLA....7.3337N}
\begin{equation}
\begin{aligned}
I_{\mathrm{bulk}}
&=\int d^2x\,\sqrt{-g}\,e^{-2\phi}\\
&\quad\times
\left[
R+4(\nabla\phi)^2
-\frac14 F_{\mu\nu}F^{\mu\nu}+Q^2
\right].
\end{aligned}
\label{action}
\end{equation}
Here $g=\det g_{\mu\nu}$ and
$F_{\mu\nu}=\partial_\mu A_\nu-\partial_\nu A_\mu$.
We consider the solution at string tree level and
to leading order in $\alpha'$.

The horizon locations satisfy
$e^{Qr_\pm}=m\pm\sqrt{m^2-q^2}$. Solving the
outer-horizon condition $g(r_+)=0$ gives
\begin{equation}
m=\tfrac12(e^{Qr_+}+q^2e^{-Qr_+}),\qquad
r_+>r_m\equiv Q^{-1}\ln|q|.
\label{domain}
\end{equation}
The extremal limit is approached as $r_+\to r_m^+$.
In this two-dimensional geometry, $r_+$ denotes the
coordinate location of the outer event horizon rather
than an areal radius. The Wald entropy increases
monotonically with $r_+$, so it provides a valid
parameterization of the admissible horizon states.

The conserved mass and electric charge, Wald entropy, and
Hawking temperature of the outer horizon are
\cite{1992NuPhB.375..421M,1992MPLA....7.3337N,
1993PhRvD..48.3427W,1994PhRvD..49.6587J,
1994PhRvD..50..846I,2001LRR.....4....6W}
\begin{equation}
\begin{aligned}
M&=2mQe^{-2\phi_0},\qquad
\mathcal Q_{el}=\sqrt2 qQe^{-2\phi_0},\\
S&=4\pi e^{-2\phi_0}\left(m+\sqrt{m^2-q^2}\right),\\
T&=\frac{Q}{2\pi}
\frac{\sqrt{m^2-q^2}}{m+\sqrt{m^2-q^2}}.
\end{aligned}\label{thermo}
\end{equation}
Here $M$ is the conserved, vacuum-subtracted energy
associated with the asymptotically normalized time
coordinate~\cite{1992NuPhB.375..421M,1996PhRvD..54.7476C,1992MPLA....7.3337N},
and $\mathcal Q_{el}=e^{-2\phi}F^{tx}$ is the conserved electric
flux. The entropy is the Wald entropy associated with
the dilaton coupling,
$S=4\pi e^{-2\phi(r_+)}$, rather than a horizon area
\cite{1993PhRvD..48.3427W,1994PhRvD..49.6587J,
1994PhRvD..50..846I,2001LRR.....4....6W}. The temperature follows from
$T=g'(r_+)/4\pi$, where the prime denotes $d/dx$,
with $m$ and $q$ held fixed.
Using Eq.~\eqref{domain}, the mass, entropy,
and temperature can be expressed in terms of $r_+$ as
\begin{equation}
\begin{aligned}
M(r_+)&=Qe^{-2\phi_0}
\left(e^{Qr_+}+q^2e^{-Qr_+}\right),\\
S(r_+)&=4\pi e^{-2\phi_0}e^{Qr_+},\\
T(r_+)&=\frac{Q}{4\pi}
\left(1-q^2e^{-2Qr_+}\right).
\end{aligned}
\label{thermo_horizon}
\end{equation}

We work in the canonical ensemble at fixed physical
electric charge $\mathcal Q_{\rm el}$ and fixed $Q$ and $\phi_0$.
Since $\mathcal Q_{\rm el}$ is proportional to $q$, fixing
the ensemble data keeps $q$ fixed as $r_+$ varies.
The electric chemical potential is
$
\mu_{\rm el}=A_t(r_+)-A_t(\infty)
=-\sqrt{2}\,q e^{-Qr_+},
$
and the first law reads
$
dM=T\,dS-\mu_{\rm el}\,d\mathcal Q_{\rm el}
$
\cite{1992MPLA....7.3337N}.
In the Euclidean convention $Z\simeq e^{-I_E}$,
the canonical on-shell action at fixed electric charge is
$I_{\mathcal Q_{\rm el}}=\beta M-S$, where $\beta=1/T$.
In these conventions,
$[M]=[\mathcal Q_{\rm el}]=[T]=[Q]=L^{-1}$ and
$[S]=[\mu_{\rm el}]=[A_t]=1$.

\section{Thermodynamic defect structure}
\label{MNY:off-shell}
\subsection{Generalized off-shell free energy}

Following the generalized off-shell free-energy
construction~\cite{2022PhRvL.129s1101W,2024PhRvD.110h1501W,2025PhRvD.111f1501W,2026SCPMA..6960401W}, we treat
$\tau>0$ as an auxiliary inverse temperature independent
of $r_+$. At fixed $(\mathcal Q_{el},Q,\phi_0)$, the charge
parameter $q$ remains fixed. The canonical off-shell free
energy is
\begin{equation}
\begin{aligned}
\mathcal F(r_+;\tau,\mathcal Q_{el})
&=M(r_+)-\frac{S(r_+)}{\tau}\\
&=Qe^{-2\phi_0}
\left[
\left(1-\frac{4\pi}{Q\tau}\right)e^{Qr_+}
+q^2e^{-Qr_+}
\right].
\end{aligned}\label{free}
\end{equation}

To construct the thermodynamic vector field, we introduce
an auxiliary angle $\Theta\in(0,\pi)$ and define
$\boldsymbol\phi=(\partial_{r_+}\mathcal F,
-\cot\Theta\csc\Theta)$
\cite{2022PhRvL.129s1101W}. Its components are
\begin{equation}
\begin{aligned}
\phi^{r_+}
&=Q^2e^{-2\phi_0}
\left[
\left(1-\frac{4\pi}{Q\tau}\right)e^{Qr_+}
-q^2e^{-Qr_+}
\right],\\
\phi^\Theta&=-\cot\Theta\csc\Theta .
\end{aligned}
\label{field}
\end{equation}
At $\Theta=0,\pi$, the angular component
$\phi^\Theta$ diverges, and the vector field points
outward at both boundaries.

With $x^\mu=(\tau,r_+,\Theta)$, the unit field
$n^a=\phi^a/\lVert\boldsymbol\phi\rVert$ is defined away
from the zeros of $\boldsymbol\phi$. Duan's
$\phi$-mapping topological current is
\cite{Duan:1984ws,1998NuPhB.514..705D,
2022PhRvL.129s1101W}
\begin{equation}
\begin{aligned}
j^\mu
&=\frac{\epsilon^{\mu\nu\rho}\epsilon_{ab}}{2\pi}
  (\partial_\nu n^a)(\partial_\rho n^b)\\
&=\delta^{(2)}(\boldsymbol\phi)
  J^\mu(\boldsymbol\phi/x).
\end{aligned}
\label{current}
\end{equation}
At fixed $\tau$, the integral of $j^\tau$ over a region
whose boundary contains no zero equals the sum of the
local winding numbers inside it. For a simple zero,
the winding number is the sign of
$J^\tau=\det[\partial(\phi^{r_+},\phi^\Theta)/
\partial(r_+,\Theta)]$.

Since $\phi^\Theta=0$ only at $\Theta=\pi/2$, the
zeros of $\boldsymbol\phi$ are determined by
$\phi^{r_+}=0$. On the nonextremal outer-horizon branch,
this gives
\begin{equation}
\tau=\beta(r_+)
=\frac{4\pi}{Q(1-q^2e^{-2Qr_+})}
=\frac{1}{T(r_+)}.
\label{zero_curve}
\end{equation}
The fixed-charge first law also gives
$\partial_{r_+}\mathcal F
=(\partial_{r_+}S)(T-1/\tau)$, with
$\partial_{r_+}S>0$.

\subsection{Defect curve, winding, and boundary behavior}

\begin{figure}[t]
\includegraphics[width=\columnwidth]{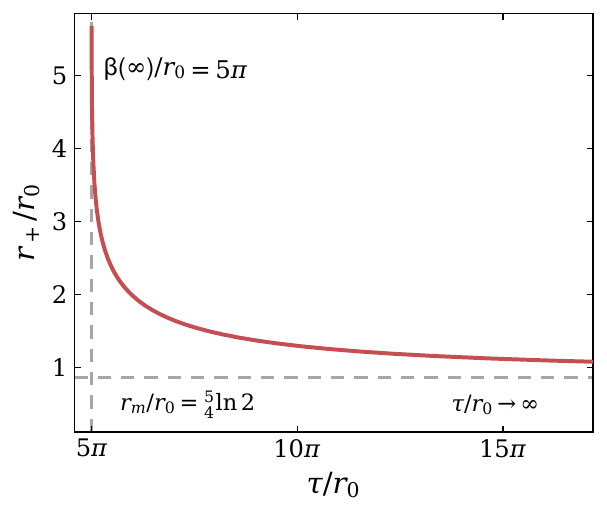}
\caption{Equilibrium defect curve for $Qr_0=4/5$ and
$q=2$. The horizontal and vertical axes are $\tau/r_0$
and $r_+/r_0$, respectively, where $r_0$ is a reference
length. The red branch is locally stable with $w=+1$.
The vertical and horizontal dashed lines mark
$\beta(\infty)/r_0=5\pi$ and
$r_m/r_0=(5/4)\ln2$, respectively.}
\label{fig:curve}
\end{figure}

On the nonextremal outer-horizon branch $r_+>r_m$,
Eq.~\eqref{zero_curve} gives
\begin{equation}
\begin{aligned}
\frac{d\beta}{dr_+}
&=-\frac{8\pi q^2e^{-2Qr_+}}
        {(1-q^2e^{-2Qr_+})^2}<0,\\
\lim_{r_+\to r_m^+}\beta(r_+)&=\infty,\qquad
\lim_{r_+\to\infty}\beta(r_+)=\frac{4\pi}{Q}.
\end{aligned}
\label{limits}
\end{equation}
The inverse temperature therefore ranges over
$(4\pi/Q,\infty)$, as shown in Fig.~\ref{fig:curve}.
For the illustrative parameters $Qr_0=4/5$ and $q=2$
used in that figure, Eqs.~\eqref{limits} and \eqref{domain}
give $\beta(\infty)/r_0=5\pi$ and
$r_m/r_0=(5/4)\ln 2$, respectively.
For each $\tau>4\pi/Q$, there is exactly one finite zero,
\begin{equation}
r_1=r_m+\frac{1}{2Q}
\ln\!\left(\frac{Q\tau}{Q\tau-4\pi}\right),
\qquad \Theta_1=\frac{\pi}{2}.
\label{root}
\end{equation}
No finite zero exists for $0<\tau\leq4\pi/Q$.
Equilibrium black holes have $0<T<Q/(4\pi)$; the
upper temperature is approached only as $r_+\to\infty$.

\begin{figure*}[htp!]
\centering
\includegraphics[width=0.33\textwidth]{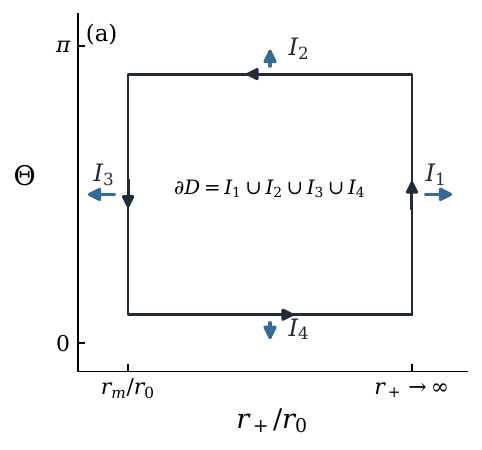}\hfill
\includegraphics[width=0.33\textwidth]{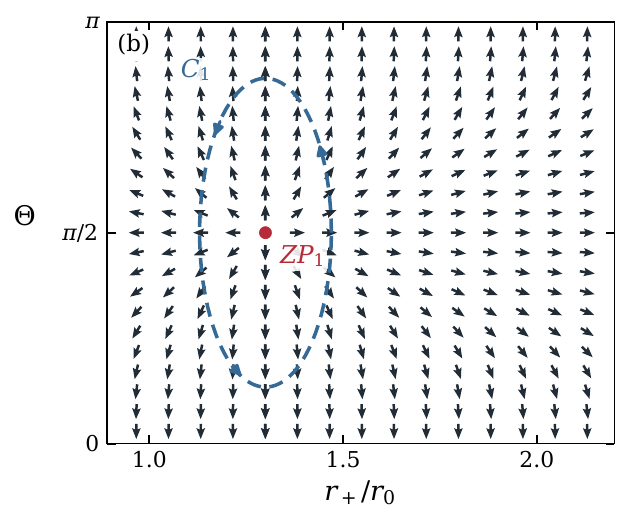}\hfill
\includegraphics[width=0.33\textwidth]{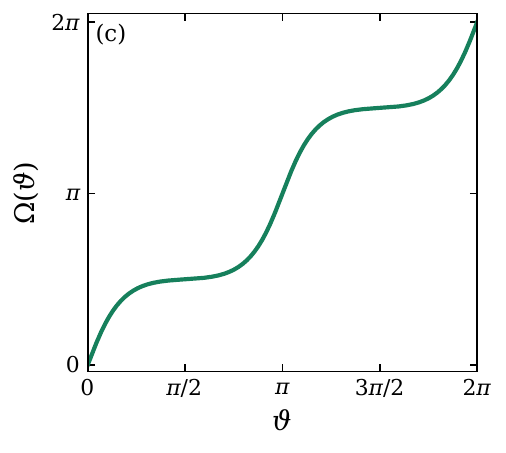}
\caption{Vector field and winding at $Qr_0=4/5$, $q=2$,
$e^{-2\phi_0}=25/16$, and $\tau/r_0=10\pi$.
(a) Rectangular counterclockwise boundary
$\partial D=I_1\cup I_2\cup I_3\cup I_4$ in
the $(r_+/r_0,\Theta)$ plane. For the numerical
regularization, its radial sides are at
$r_+/r_0=(5/4)\ln2+0.10$ and $2.50$, and
its angular sides at $\Theta=0.30$ and
$\pi-0.30$. Black arrows show contour orientation;
blue arrows show the vector-field directions.
The horizontal labels represent the physical
limits $r_+\to r_m^+$ and $r_+\to\infty$;
neither endpoint lies on the numerical contour.
(b) Normalized vector-field plot, with the sole zero
$ZP_1$ and the updated local contour $C_1$
of Eq.~\eqref{eq:C1param}.
(c) Accumulated deflection along $C_1$,
with $\Omega(2\pi)=2\pi$ and $w_1=+1$.
Since this is the only zero, $W=+1$.}
\label{fig:vector}
\end{figure*}

At the zero $(r_1,\pi/2)$, the mixed derivatives vanish,
while
\begin{equation}
\begin{aligned}
\left.\partial_{r_+}\phi^{r_+}\right|_1
&=2Q^3q^2e^{-2\phi_0}e^{-Qr_1}>0,\\
\left.\partial_\Theta\phi^\Theta\right|_1&=1.
\end{aligned}
\end{equation}
Thus $J_1>0$. For the counterclockwise contour $C_1$
shown in Fig.~\ref{fig:vector}, the winding number is
\begin{equation}
w_1=\frac{1}{2\pi}\oint_{C_1}
d\arg(\phi^{r_+}+i\phi^\Theta)
=\operatorname{sgn}J_1=+1.
\end{equation}
Since $\partial_{r_+}^2\mathcal F|_1>0$ at fixed
$\mathcal Q_{el}$ and $\tau$, this equilibrium branch is locally
stable under variations of $r_+$. It is the only
finite-horizon branch, so
\begin{equation}
W=\sum_i w_i=w_1=+1
\qquad(\tau>4\pi/Q).
\label{finitecount}
\end{equation}

For Fig.~\ref{fig:vector}, we use $Qr_0=4/5$, $q=2$,
$e^{-2\phi_0}=25/16$, and $\tau/r_0=10\pi$.
Writing $u=r_+/r_0$ and $\hat\tau=\tau/r_0$,
the components used for plotting are obtained directly
from Eq.~\eqref{field}:
\begin{equation}
\begin{aligned}
\widetilde\phi^{r_+}
&=r_0^2\phi^{r_+}\\
&=(Qr_0)^2e^{-2\phi_0}
\Biggl[
\left(1-\frac{4\pi}{(Qr_0)\hat\tau}\right)
e^{(Qr_0)u}\\
&\hspace{3em}
-q^2e^{-(Qr_0)u}
\Biggr],\\
\widetilde\phi^\Theta
&=\phi^\Theta
=-\cot\Theta\csc\Theta .
\end{aligned}
\label{eq:plotfield}
\end{equation}
Here $(Qr_0)^2e^{-2\phi_0}=1$. The only zero
at this fixed auxiliary inverse temperature is
$(u_1,\Theta_1)=(15\ln2/8,\pi/2)$.

The local counterclockwise contour $C_1$ in
Fig.~\ref{fig:vector}(b) is parametrized by
\begin{equation}
u(\vartheta)=u_1+0.17\cos\vartheta,
\Theta(\vartheta)=\frac{\pi}{2}
                 +1.15\sin\vartheta,
 0\leq\vartheta\leq2\pi.
\label{eq:C1param}
\end{equation}
Following the contour-deflection method
\cite{2022PhRvL.129s1101W,2024EPJC...84.1204H},
we set $\widetilde n^a=
\widetilde\phi^a/|\widetilde{\boldsymbol\phi}|$ and
define the continuously accumulated angle
\begin{equation}
\Omega(\vartheta)=\int_0^\vartheta
\epsilon_{ab}\widetilde n^a(\vartheta')
\frac{d\widetilde n^b(\vartheta')}{d\vartheta'}
\,d\vartheta',\qquad
w_1=\frac{\Omega(2\pi)}{2\pi},
\label{eq:deflection}
\end{equation}
where $\epsilon_{r_+\Theta}=+1$ and $\Omega(0)=0$.
The positive rescaling of the radial component in
Eq.~\eqref{eq:plotfield} preserves the zeros and their
winding numbers, although the detailed profile of
$\Omega(\vartheta)$ depends on this plotting normalization
and the choice of $C_1$. Along the contour specified in
Eq.~\eqref{eq:C1param}, Fig.~\ref{fig:vector}(c) gives
$\Omega(2\pi)=2\pi$, and hence $w_1=+1$, in agreement
with the Jacobian result. 
Following Ref.~\cite{2026SCPMA..6960401W}, the
counterclockwise boundary in Fig.~\ref{fig:vector}(a)
is $\partial D=I_1\cup I_2\cup I_3\cup I_4$, where
$I_1=\{u=u_{\mathrm R},\,
\Theta:\epsilon\to\pi-\epsilon\}$,
$I_2=\{\Theta=\pi-\epsilon,\,
u:u_{\mathrm R}\to u_{\mathrm L}\}$,
$I_3=\{u=u_{\mathrm L},\,
\Theta:\pi-\epsilon\to\epsilon\}$, and
$I_4=\{\Theta=\epsilon,\,
u:u_{\mathrm L}\to u_{\mathrm R}\}$.
Here $u_{\mathrm L}=(5/4)\ln2+0.10$,
$u_{\mathrm R}=2.50$, and $\epsilon=0.30$.
The outward normal field directions on these four
segments are, respectively,
$(\rightarrow,\uparrow,\leftarrow,\downarrow)$.
The boundary encloses only $ZP_1$ and contains no zero;
thus its degree is
\begin{equation}
\deg(\widetilde{\boldsymbol\phi},D,0)
=\frac{1}{2\pi}\oint_{\partial D}
d\arg\!\left(\widetilde\phi^{r_+}
+i\widetilde\phi^\Theta\right)
=w_1=W=+1 .
\end{equation}

The zero-point curve shown in Fig.~\ref{fig:curve}
therefore corresponds to a thermodynamic phase structure
consisting solely of a locally stable black-hole branch
with positive winding number ($w=+1$).
As $\tau$ decreases from infinity to $(4\pi/Q)^+$,
the corresponding zero point moves continuously from
$r_+\to r_m^+$ to $r_+\to\infty$.
For any finite $R>r_m$, the zero point crosses $r_+=R$
at $\tau_R=\beta(R)>4\pi/Q$, and
$\tau_R\to4\pi/Q$ as $R\to\infty$.
At $\tau=4\pi/Q$, the vector field approaches zero
as $r_+\to\infty$ along $\Theta=\pi/2$.
Thus, the disappearance of the stable black hole branch
is associated with the large-radius limit.
No generation or annihilation point occurs at finite $r_+$,
since the defect curve is strictly monotonic.

\section{Thermodynamic topological classification}
\label{MNY:topo_class}

\begin{table*}[ht!]
\caption{Comparison of thermodynamic topological (sub)classes
with the charged MNY black hole.
The patterns $W^{0+\leftrightarrow1-}$ and
$\widehat W^{1-}$ remain candidates for possible new
thermodynamic topological (sub)classes whose physical
realizations have yet to be established (see Ref.~\cite{2025PhRvD.111f1501W}).
$\mathrm S$ and $\mathrm U$ denote locally stable and
unstable black hole branches; the subscripts ${\rm s}$ and $\ell$
denote small and large black holes. $\varnothing$ denotes
no equilibrium black hole. Innermost and outermost refer
to defect branches ordered by event-horizon size.
Here, $r_m$ denotes the minimal possible event horizon radius
of the black hole, which may be zero or nonzero; $\beta_{\mathrm f}\in(0,\infty)$ denotes a finite
positive limiting value of the inverse temperature,
termed ``fixed temperature'' in
Ref.~\cite{2025PhRvD.111f1501W};
its value depends on the black-hole solution and parameters.
}
\label{tab:classes}
\begingroup\footnotesize\setlength{\tabcolsep}{2.2pt}
\begin{ruledtabular}
\begin{tabular}{lccccccc}
Class/subclass & $(\beta(r_m),\beta(\infty))$ &
Innermost & Outermost & Low $T$ ($\tau\to\infty$) & High $T$ ($\tau\to0$) & DP & $W$\\
\hline
$W^{1-}$~\cite{2024PhRvD.110h1501W} & $(0,\infty)$ & $\mathrm U$ & $\mathrm U$ & $\mathrm U_\ell$ & $\mathrm U_{\rm s}$ & In pairs & $-1$\\
$W^{0+}$~\cite{2024PhRvD.110h1501W} & $(\infty,\infty)$ & $\mathrm S$ & $\mathrm U$ & $\mathrm S_{\rm s}+\mathrm U_\ell$ & $\varnothing$ & One More GP & $0$\\
$W^{0-}$~\cite{2024PhRvD.110h1501W} & $(0,0)$ & $\mathrm U$ & $\mathrm S$ & $\varnothing$ & $\mathrm U_{\rm s}+\mathrm S_\ell$ & One More AP & $0$\\
$W^{1+}$~\cite{2024PhRvD.110h1501W} & $(\infty,0)$ & $\mathrm S$ & $\mathrm S$ & $\mathrm S_{\rm s}$ & $\mathrm S_\ell$ & In pairs & $+1$\\
$W^{0-\leftrightarrow1+}$~\cite{2025PhRvD.111f1501W} & $(\beta_{\mathrm f},0)$ & $\mathrm U$ & $\mathrm S$ & $\varnothing$ & $\mathrm S_\ell$ & One More AP & $0,+1$\\
$\overline W^{1+}$~\cite{2025PhRvD.111f1501W} & $(\beta_{\mathrm f},0)$ & $\mathrm S$ & $\mathrm S$ & $\varnothing$ & $\mathrm S_\ell$ & In pairs & $+1$\\
$\widehat W^{1+}$~\cite{2025PhRvD.111f1501W} & $(\infty,0)$ & $\mathrm S$ & $\mathrm S$ & $\mathrm U_{\rm s}+2\mathrm S_{\rm s}$ & $\mathrm S_\ell$ & One More GP & $+1$\\
$\widetilde W^{1+}$~\cite{2025PhRvD.112l4024A} & $(0,0)$ & $\mathrm U$ & $\mathrm S$ & $\mathrm S_{\rm s}$ & $\mathrm U_{\rm s}+\mathrm S_{\rm s}+\mathrm S_\ell$ & One More AP & $+1$\\
$\ddot W^{1-}$~\cite{2025EPJC...85.1386C} & $(0,\infty)$ & $\mathrm U$ & $\mathrm S$ & $\mathrm U_{\rm s}$ & $2\mathrm U_{\rm s}+\mathrm S_\ell$ & One More AP & $-1$\\
$\overline W^{0-}$~\cite{2026EPJC...86..187W} & $(0,0)$ & $\mathrm U$ & $\mathrm S$ & $\mathrm U_{\rm s}+\mathrm S_{\rm s}$ & $\mathrm U_{\rm s}+\mathrm S_\ell$ & In pairs & $0$\\
$\widetilde W^{0-}$~\cite{2026arXiv260714173W} & $(0,0)$ & $\mathrm U$ & $\mathrm S$ & $\mathrm S_{\rm s}$ & $\mathrm U_{\rm s}+\mathrm S_\ell$ & In pairs & $0$\\
$\overline W^{1-}$~\cite{2026arXiv260916568Z,2026arXiv260920898Y} & $(\beta_{\mathrm f},\infty)$ & $\mathrm U$ & $\mathrm U$ & $\mathrm U_\ell$ & $\varnothing$ & In pairs & $-1$\\
$W^{0+\leftrightarrow1-}$\,~\cite{2025PhRvD.111f1501W} & $(\beta_{\mathrm f},\infty)$ & $\mathrm S$ & $\mathrm U$ & $\mathrm U_\ell$ & $\varnothing$ & One More GP & $0,-1$\\
$\widehat W^{1-}$\,~\cite{2025PhRvD.111f1501W} & $(0,\infty)$ & $\mathrm U$ & $\mathrm U$ & $\mathrm U_\ell$ & $\mathrm S_{\rm s}+2\mathrm U_{\rm s}$ & One More AP & $-1$\\
\textbf{$\breve W^{1+}$} & $(\infty,\beta_{\mathrm f})$ & $\mathrm S$ & $\mathrm S$ & $\mathrm S_{\rm s}$ & $\varnothing$ & In pairs & $+1$
\end{tabular}
\end{ruledtabular}
\endgroup
\end{table*}

The MNY black hole has a single stable branch with
$W=+1$, but its inverse temperature approaches a
finite positive value as $r_+\to\infty$.
Table~\ref{tab:classes} compares this behavior with
previously reported thermodynamic topological (sub)classes.
%Here $\beta_{\mathrm f}$ denotes the finite positive endpoint value termed ``fixed temperature'' in Ref.~\cite{2025PhRvD.111f1501W}. Its value may differ between black-hole solutions; for MNY, $\beta_{\mathrm f}=4\pi/Q$.
Following Ref.~\cite{2025PhRvD.111f1501W},
``In pairs'' means that finite generation points (GPs)
and annihilation points (APs), if present, occur in
equal numbers; the zero-pair case is included.
The MNY solution realizes the zero-pair case.

For MNY, the single stable branch is both the innermost
and outermost black-hole state. At low temperature it
approaches the extremal bound, as in the $W^{1+}$ class.
At large horizon radius, however, its inverse temperature
approaches $4\pi/Q$, whereas $\beta(\infty)=0$ for
$W^{1+}$~\cite{2024PhRvD.110h1501W}. Thus both have
$W=+1$ and stable limiting branches, but their
large-horizon asymptotic behaviors differ.

The $W^{0-\leftrightarrow1+}$,
$\overline W^{1+}$, $W^{0+\leftrightarrow1-}$,
and $\overline W^{1-}$ (sub)classes have a finite
inverse-temperature limit at the lower horizon boundary,
but $\beta(\infty)$ is either zero or infinite
\cite{2025PhRvD.111f1501W,2026arXiv260920898Y}.
The remaining (sub)classes in Table~\ref{tab:classes}
also have $\beta(\infty)=0$ or $\infty$. None exhibits the finite, nonzero large-horizon
asymptotic behavior of $\beta(r_+)$ found for MNY.

This finite limit follows from the common exponential
growth of the mass and dilaton entropy:
$(dS/dM)_{\mathcal Q_{el}}=\beta(r_+)\to4\pi/Q$ as
$r_+\to\infty$. The stable equilibrium branch therefore
reaches the large-horizon boundary at a finite auxiliary
inverse temperature, leaving no finite-horizon state at
higher temperatures. This distinct defect behavior,
despite $W=+1$, motivates a new thermodynamic
topological subclass.

We characterize this new subclass by
\begin{equation}
\breve W^{1+}:\qquad
\bigl(\beta(r_m),\beta(\infty)\bigr)
=(\infty,\beta_{\mathrm f}),\qquad
\beta_{\mathrm f}>0.
\label{newclass}
\end{equation}
This differs from standard \(W^{1+}\), for which \(\beta(\infty) = 0\).
For the charged MNY solution, $\beta_{\mathrm f}=4\pi/Q$
and $[w_i]=[+]$, with $W=+1$ for
$\tau>\beta_{\mathrm f}$.
To examine whether other black holes exhibit the same thermodynamic topological subclass, we analyze a four-dimensional uncharged Ho\v{r}ava--Lifshitz black hole in Appendix~\ref{app:HL}, focusing on its inverse-temperature asymptotics.

Following the symmetry-based proposals of candidate
thermodynamic topological subclasses in
Refs.~\cite{2025PhRvD.111f1501W,2025PhRvD.112l4024A},
the finite large-horizon asymptotic behavior found here suggests a
possible counterpart with $\beta(r_m)=0$ and
$\beta(\infty)=\beta_{\mathrm f}>0$. If realized with
negative winding, this pattern would define a candidate
$\breve W^{1-}$ subclass. Whether this asymptotic behavior is realized by a black-hole solution remains to be established.

More generally, the equilibrium condition derived from
the generalized off-shell free energy depends on the radial derivatives
of the mass and entropy, so changing the entropy
prescription can alter the defect curve and its
thermodynamic topological classification.
Generalized entropy prescriptions
(e.g., Rényi, Barrow, Kaniadakis, and Tsallis entropies)
have been shown to modify thermodynamic phase structures
and thermodynamic topological classifications
\cite{2024EPJC...84..826T,
2025EPJC...85..305G,
2025PDU....4701833G,
2025PhLB..86839720L,
2024arXiv240808325H}.
These studies motivate further investigation of
whether entropy modifications can produce finite, nonzero
large-horizon limits of the corresponding equilibrium inverse temperature in other black-hole families, accompanied by distinct defect-branch
structures and thermodynamic topological (sub)classes.

\section{Conclusions}
\label{MNY:conclusion}

We have investigated the thermodynamic topology of the charged MNY black hole in the fixed-charge ensemble. For $\tau>4\pi/Q$, the vector field has a single zero point with winding number $+1$, corresponding to a stable black-hole state. As $\tau$ decreases toward $4\pi/Q$, its horizon radius tends to infinity. No finite-horizon zero point exists at or below this value, and no generation or annihilation point occurs at finite radius.

The finite large-horizon limit follows from the asymptotic mass--entropy relation, which gives $\beta=(dS/dM)_{\mathcal Q_{el}}\to4\pi/Q$. The MNY black hole therefore differs from the standard $W^{1+}$ class despite having the same topological number and stable innermost and outermost states. In particular, the standard class contains a stable large black hole in the high-temperature limit, whereas the MNY black hole has no equilibrium state in that limit. We denote this distinct subclass by $\breve W^{1+}$.

This example shows that the topological number and the stability of the limiting states do not suffice to distinguish black holes with different high-temperature behavior. The large-horizon asymptotic behavior of the inverse temperature is also relevant to their thermodynamic topological classification. 
Our findings broaden the thermodynamic topological classification of black holes by including a finite, nonzero large-horizon inverse-temperature limit and suggest that further (sub)classes may be hidden in other black-hole families.

\section*{Acknowledgments}
This work is supported by Guizhou provincial natural science foundation project No. ZD[2026]058.

\appendix
\section{A Ho\v{r}ava--Lifshitz example}
\label{app:HL}

As a further example, we consider the Ho\v{r}ava--Lifshitz black holes studied in Refs.~\cite{2009PhLB..679..504C,2011PhRvD..83f4015C}. The metric is
\begin{equation}
\begin{aligned}
ds^2&=-N^2(r)f(r)\,dt^2
      +\frac{dr^2}{f(r)}+r^2d\Omega_k^2,\\
f(r)&=k-\Lambda r^2-\alpha r^s,\qquad
N(r)=\gamma r^{1-2s},
\end{aligned}
\label{eq:HL-general-metric}
\end{equation}
where \(\Lambda<0\), \(\alpha\) and \(\gamma\) are integration constants, and
$
s=\frac{2\lambda-\sqrt{2(3\lambda-1)}}{\lambda-1}
$
on the negative-sign solution branch. The event horizon \(r_+\) is the largest positive root of \(f(r)=0\).

The mass, entropy, and Hawking temperature are \cite{2009PhLB..679..504C,2011PhRvD..83f4015C}
\begin{equation}
\begin{aligned}
M&=\frac{c^3\gamma\Omega_k}{16\pi G\Lambda}
   \frac{1+s}{s-2}
   \frac{(k-\Lambda r_+^2)^2}{r_+^{2s}},\\
S&=\frac{c^3\Omega_k}{4G\Lambda}
   \frac{1+s}{s-2}
   \left[k\ln(-\Lambda r_+^2)-\Lambda r_+^2\right]
   +S_0,\\
T&=\frac{\gamma}{4\pi r_+^{2s}}
   \left[-\Lambda r_+^2(2-s)-sk\right],
\end{aligned}
\label{eq:HL-general-thermo}
\end{equation}
where \(\Omega_k\) is the volume of the unit horizon and \(S_0\) is an entropy integration constant. At fixed theory couplings and time normalization, these quantities satisfy \(dM=T\,dS\). For the generalized off-shell free energy \(\mathcal F=M-S/\tau\), the radial component of the thermodynamic vector field is
$
\phi^{r_+}=S'(r_+)\left(T-\frac{1}{\tau}\right).
$
The equilibrium condition gives
\begin{equation}
\tau=
\frac{4\pi r_+^{2s}}
{\gamma[-2\Lambda r_+^2-s(k-\Lambda r_+^2)]},
\qquad \Theta=\frac{\pi}{2}.
\label{eq:HL-general-zero}
\end{equation}

We consider \(\lambda=3\), \(s=1\), and \(k=1\), as in Fig.~5 of Ref.~\cite{2024EPJC...84...96C}, where the defect curve and its positive winding number were analyzed. Here we focus on the finite large-horizon inverse-temperature limit and its interpretation within the \(\breve W^{1+}\) subclass. Setting \(\Lambda=-\ell^{-2}\), \(\gamma=\ell\), and \(c=1\), we obtain
$
N(r)=\frac{\ell}{r},\qquad
f(r)=1+\frac{r^2}{\ell^2}-\alpha r,
$
with \(g_{tt}\to-1\) as \(r\to\infty\). The horizon relation \(r_+r_-=\ell^2\) implies that the nonextremal outer-horizon branch has \(r_+>r_m=\ell\).

For this branch, the inverse-temperature defect curve is
$
\tau=\beta(r_+)=\frac{4\pi\ell}{1-\ell^2/r_+^2},
$
with
$
\beta(r_m^+)=\infty,\qquad
\beta(\infty)=\beta_{\mathrm f}=4\pi\ell>0.
$
A unique finite zero therefore exists for \(\tau>\beta_{\mathrm f}\), at
$
r_1=\frac{\ell}{\sqrt{1-\beta_{\mathrm f}/\tau}},
\qquad \Theta_1=\frac{\pi}{2}.
$
Since \(J_1=S'(r_1)T'(r_1)>0\), its winding number is \(w_1=+1\), giving \(W=+1\). No finite-horizon zero exists for \(0<\tau\leq\beta_{\mathrm f}\). Thus, this solution exhibits the same inverse-temperature limiting behavior and single locally stable branch as the charged MNY black hole.

\bibliography{references}

\end{document}